\documentclass[aps,prb,twocolumn,floats]{revtex4}

\usepackage{amssymb}
\usepackage{amsmath}
\usepackage{graphicx}
\usepackage{subfigure}
\usepackage{bm}
\usepackage{mdframed}
\usepackage[pangram]{blindtext}

\begin{document}

\title{Spin relaxation in corrugated graphene}

\author{I. M. Vicent,$^1$ H. Ochoa,$^{2}$ and F. Guinea,$^{1,3}$}

\affiliation{$^1$Instituto Madrile\~no de Estudios Avanzados en Nanociencia (IMDEA-Nanociencia), 28049 Madrid, Spain\\
$^2$ Department of Physics and Astronomy, University of California, Los Angeles, California 90095, USA
\\
$^3$Department of Physics and Astronomy, University of Manchester, Manchester M13 9PL, UK}

%\date{\today}

\begin{abstract}
In graphene, out-of-plane (flexural) vibrations and static ripples imposed by the substrate relax the electron spin, intrinsically protected by mirror symmetry. We calculate the relaxation times in different scenarios, accounting for all the possible spin-phonon couplings allowed by the hexagonal symmetry of the lattice. Scattering by flexural phonons imposes the ultimate bound to the spin lifetimes, in the ballpark of hundreds of nano-seconds at room temperature. This estimate and the behavior as a function of the carrier concentration are substantially altered by the presence of tensions or the pinning with the substrate. Static ripples also influence the spin transport in the diffusive regime, dominated by motional narrowing. We find that the D'yakonov-Perel' mechanism saturates when the mean free path is comparable to the correlation length of the heights profile. In this regime, the spin-relaxation times are exclusively determined by the geometry of the corrugations. Simple models for typical corrugations lead to lifetimes  of the order of tens of micro-seconds.
\end{abstract}
%\pacs{}

\maketitle

\textit{Introduction}.---Since the injection and detection of spin currents was experimentally demonstrated,\cite{tombros} graphene is considered as a very appealing element in spintronics\cite{dasarma} devices. The spin polarization of the currents is expected to survive over long distances due to the weakness of the spin-orbit coupling\cite{4} and almost complete absence of nuclear magnetic moments. However, experimental studies yield spin diffusion lengths several orders of magnitude shorter\cite{7,6,8} than early theoretical predictions.\cite{22,11,12} Recent years have witnessed a fast development of the field. In the theoretical side, new models of spin relaxation have been proposed,\cite{kawakami_fabian,natPhys_roche,fabian_resonant,cazalilla_resonant} whereas the experimental efforts have been focused on the efficiency of spin injection\cite{kawakami_spin_injection,Fert_spin_injection} and the isolation of the samples from the environment.\cite{VanWees,vanWees_gr/BN,NanoLett2016}

The spin-relaxation processes in graphene involve inter-band transitions between states of opposite parity with respect to mirror ($z\rightarrow -z$) reflection, which make them intrinsically weak. These processes can be assisted by disorder in some cases; for example, resonant impurities induce a local sp$^3$-like distortion of the lattice, hybridizing $\pi$ and $\sigma$ electronic states.\cite{castro-neto_guinea} This is a particular example of the Elliot-Yafet mechanism,\cite{EY1} in which the spin-relaxation times are proportional to the elastic scattering times dominating charge transport. This contrasts with the D'yakonov-Perel' mechanism,\cite{DP} in which this relation is inverse due to a motional narrowing process. The interplay between charge and spin diffusion in graphene has been an object of debate since the first studies in this material.\cite{6,8,EY2}

Corrugations and thermal vibrations in the out-of-plane direction, on the other hand, break explicitly the mirror symmetry, mixing electronic states with opposite parity. In this Letter, we evaluate the spin lifetimes limited by this unavoidable source of relaxation. Our analysis contains all the possible spin-lattice couplings allowed by symmetry in weakly corrugated graphene layers. We find that the scattering with flexural phonons limits the spin-relaxation times down to $\tau_s\sim 100$ $ns$ in suspended samples. We also discuss the deviation from the usual D'yakonov-Perel' mechanism in the diffusive regime, of relevance in epitaxial graphene.

\textit{Spin-lattice coupling}.---We consider the low-energy description of graphene $\pi$-electrons around the two inequivalent corners of the hexagonal Brillouin zone, $\mathbf{K}_{\pm}$. The Hamiltonian reads as $\mathcal{H}=\hbar\, v_F\, \boldsymbol{\Sigma}\cdot\mathbf{k} +\mathcal{H}_{SO}$, with $v_F\approx10^6$ m/s. The first term describes the Dirac bands, where the operators $\boldsymbol{\Sigma}=\left(\pm\sigma_x,\sigma_y\right)$ are Pauli matrices acting on the sub-lattice degrees of freedom of the spinor wave function. The second term accounts for relativistic (spin-orbit) effects. In corrugated samples, it can be generically written as $\mathcal{H}_{SO}=
\pm\,\Delta\,\sigma_z\,s_z\,+\,\mathcal{H}_{\textrm{s-l}}$, 
where the first term is the Kane-Mele coupling,\cite{KM} $s_i$ being Pauli matrices associated with the spin degree of freedom; the strength of this coupling is of the order of $\mu$eV,\cite{4} so it will be neglected from now on. The second term represents the coupling between the electron spin and the lattice degrees of freedom due to the breakdown of the mirror symmetry. These couplings appear as invariants of the $C_{6v}$ point group symmetry of the lattice; the most generic Hamiltonian reads\cite{8a}
\begin{widetext}
\begin{align}
\label{eq:Hsl}
\mathcal{H}_{\textrm{s-l}}=\beta_{BR} \left( \bm{\Sigma} \times \bm{s}\right)_z \nabla^2 h +
\beta_D \left[\left( \bar{\bm{\Sigma}} \times \bm{s}\right)_z
 \left(\partial_{y}^2-\partial_{x}^2\right)h+
 2\,\bar{\bm{\Sigma}}\cdot\bm{s}\,\partial_{x}\partial_{y} h\right]\pm
\beta_{\lambda} \left[2\, \partial_x\partial_yh\, s_x + (\partial_y^2-\partial_x^2)h\, s_y\right],
\end{align}
\end{widetext}
where $h\left(\mathbf{x}\right)$ is the height profile and the bar stands for complex conjugation, $\bar{\bm{\Sigma}}=\left(\pm\sigma_x,-\sigma_y\right)$. The first two terms can be interpreted as spin-dependent hopping processes resulting from virtual transitions into the $\sigma$-bands. The first one acquires the form of the usual Bychkov-Rashba coupling,\cite{BR} whereas the second term resembles the form of a Dresselhaus coupling.\cite{D} The last term in Eq.~\eqref{eq:Hsl} can be understood as a spin-dependent correction to the crystal field. A tight-binding calculation\cite{8a} gives  (in units of $\hbar\, v_F$) $\beta_{BR,\lambda}\sim 5 \cdot 10^{-4}$; $\beta_{D}$ is much weaker, it appears only when considering hoppings beyond nearest neighbors.

The spin-phonon coupling can be derived from Eq.~\eqref{eq:Hsl} by promoting the height profile to a dynamical variable. Following the standard quantization procedure, we identify the Fourier components of the out-of-plane displacements with the flexural phonon operators as\begin{align}
h\left(\mathbf{q}\right)\longrightarrow\sqrt{\frac{\hbar}{2\rho\,\omega_{\mathbf{q}}}}\left[d_{\mathbf{q}}+
\left(d_{-\mathbf{q}}\right)^{\dagger}\right],
\end{align}
where  $\rho\approx 7.6 \cdot 10^{-7}$ kg m$^{-2}$  is the carbon-mass density. We consider only long-wavelength modes, so we neglect inter-valley scattering and the contribution from the optical branch. The dispersion relation can be written as\cite{Vivek,Bruno}\begin{align}
\omega_{\mathbf{q}}=\sqrt{\frac{\kappa}{\rho}}\times\sqrt{\left|\mathbf{q}\right|^4+\frac{\vartheta_{ij}}{\kappa}q_iq_j+\gamma^4}.
\end{align}
The anharmonic coupling with the in-plane modes linearizes the dispersion relation at low momenta, introducing a cut-off\cite{foot_anhar} in the quadratic dispersion of the bending modes. In this expression $\kappa\approx0.8$ eV represents the bending rigidity\cite{8a} of the graphene membrane. Tensions breaking the full-rotational symmetry produce the same effect. For simplicity, we consider the case of an isotropic tension of the form $\vartheta=Ku$, where $K\approx 21$ eV $\mathrm{\AA}^{-2}$ is the 2D bulk modulus\cite{modulus} and $u$ is the strain of the lattice; we define then $q_c=\sqrt{Ku/\kappa}$. In supported samples, the interaction with the substrate introduces an additional momentum scale $\gamma\approx 0.1 \, \mathrm{\AA^{-1}}$ related to the pinning lengths.\cite{pinning}

\begin{figure}
\subfigure{\includegraphics[width=0.48\columnwidth]{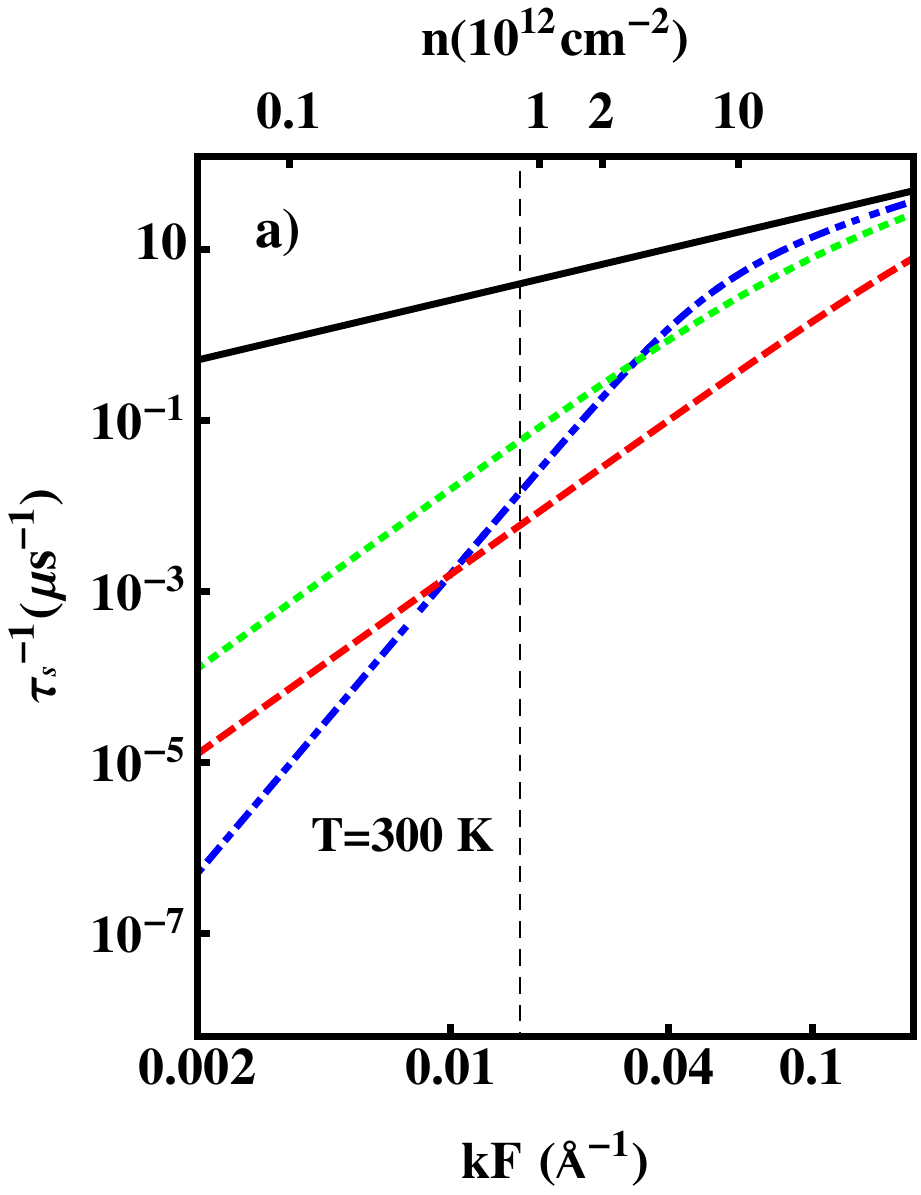}}
\subfigure{\includegraphics[width=0.48\columnwidth]{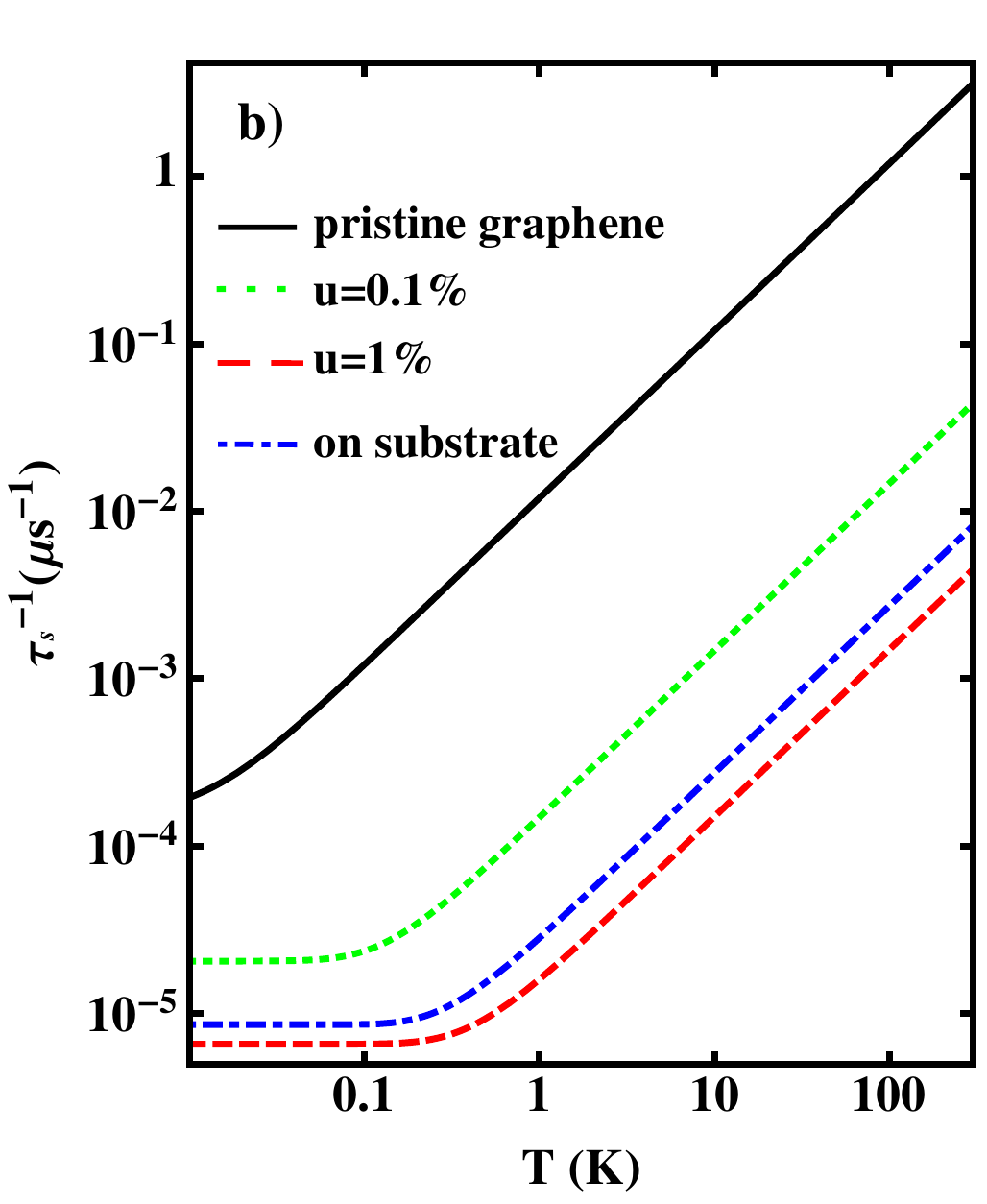}}
\caption{ Spin relaxation rates due to the scattering by flexural phonons as a function of a) carrier concentration (where $T=300$ K) and b) temperature (where $\epsilon_F=0.1$ eV, corresponding to the vertical dashed line in a).}
\label{fig:tau1}
\end{figure}

\textit{Spin relaxation due to flexural phonons}.---We consider first the spin lifetimes limited by electron-phonon scattering in the absence of other sources of disorder. In the spirit of Matthiessen's rule, the spin lifetimes limited by each of the couplings separately are combined in a single relaxation rate. A Fermi's golden rule calculation gives\begin{align}
\label{eq:Fermi_gr}
\frac{1}{\tau_s}=\frac{1}{\pi\hbar^2v_F}\int_0^{2k_F}dq\,\frac{\left|\hat{\Gamma}_{k_F,q}\right|^2}{\sqrt{1-\left(\frac{q}{2k_F}\right)^2}}
\left(2n_{q}+1\right),
\end{align}
where $n_{\mathbf{q}}=\left(e^{\hbar\omega_{q}/k_BT}-1\right)^{-1}$ and the squared matrix elements of the electron-phonon coupling read\begin{align}
\left|\hat{\Gamma}_{k_F,q}\right|^2=\frac{\hbar q^4}{2\rho\,\omega_q}\left[\beta_{BR}^2+\beta_{D}^2+\beta_{\lambda}^2\left(1-\frac{q^2}{4k_F^2}\right)\right].
\end{align}
In the derivation of Eq.~\eqref{eq:Fermi_gr} we have employed a quasi-elastic approximation --i.e., we have neglected the phonon contribution in the energy-conservation constrain-- 
provided that $T\ll T_F\equiv\epsilon_F/k_B$ for the usual dopings, where $\epsilon_F=\hbar v_F k_F$ is the Fermi energy measured with respect the Dirac point. Notice also that the $\beta_{\lambda}$-coupling preserves the chirality of the wave function, so this channel is absent under the backscattering condition, $q=2k_F$.

The spin-relaxation rates evaluated from Eq.~\eqref{eq:Fermi_gr} are shown in Fig.~\ref{fig:tau1}. In the free-standing case (black continuous curve) the spin lifetimes are limited to a few hundreds of nano-seconds. Tensions (green dotted and red dashed curves) and the interaction with the substrate (blue dashed-dotted curve) suppress the contribution from flexural modes at the lowest momenta, modifying also the dependence on the carrier concentration as it is shown in panel a. The two different regimes shown in panel b are determined by the Bloch-Gr\"uneisen temperature, $T_{BG}=\hbar\, \omega_{2k_F}/k_B$, of hundreds of mK at most. In the experimentally most relevant regime, $T\gg T_{BG}$, the spin-relaxation rates in suspended samples are given by the expression\begin{align}
\frac{1}{\tau_s}\approx\frac{\tilde{\beta}^2k_F}{2\,\hbar^2v_F}\frac{k_BT}{\kappa}\left(\frac{2k_F}{q_c}\right)^{2\nu},
\end{align}
where $\nu=0,1$ corresponds to the free-standing ($q_c\ll 2k_F$) and strained cases, respectively. The effective spin-phonon coupling reads $\tilde{\beta}^2\equiv\beta_{BR}^2+\beta_{D}^2+\beta_{\lambda}^2/4$. In supported samples, the pinning effects become relevant at $\gamma>2k_F$, for which\begin{align}
\frac{1}{\tau_s}\approx\frac{2\,\tilde{\beta}^2\,k_F^5}{\hbar^2\,v_F\,\gamma^4}\frac{k_BT}{\kappa}.
\end{align}

At low temperatures, $T\ll T_{BG}$, the spin-relaxation rates behave as $\sim T^{3/2}$ ($T^{4}$) in the free-standing (strained) case, whereas they are exponentially suppressed in pinned samples.

\begin{figure}
\includegraphics[width=1\columnwidth]{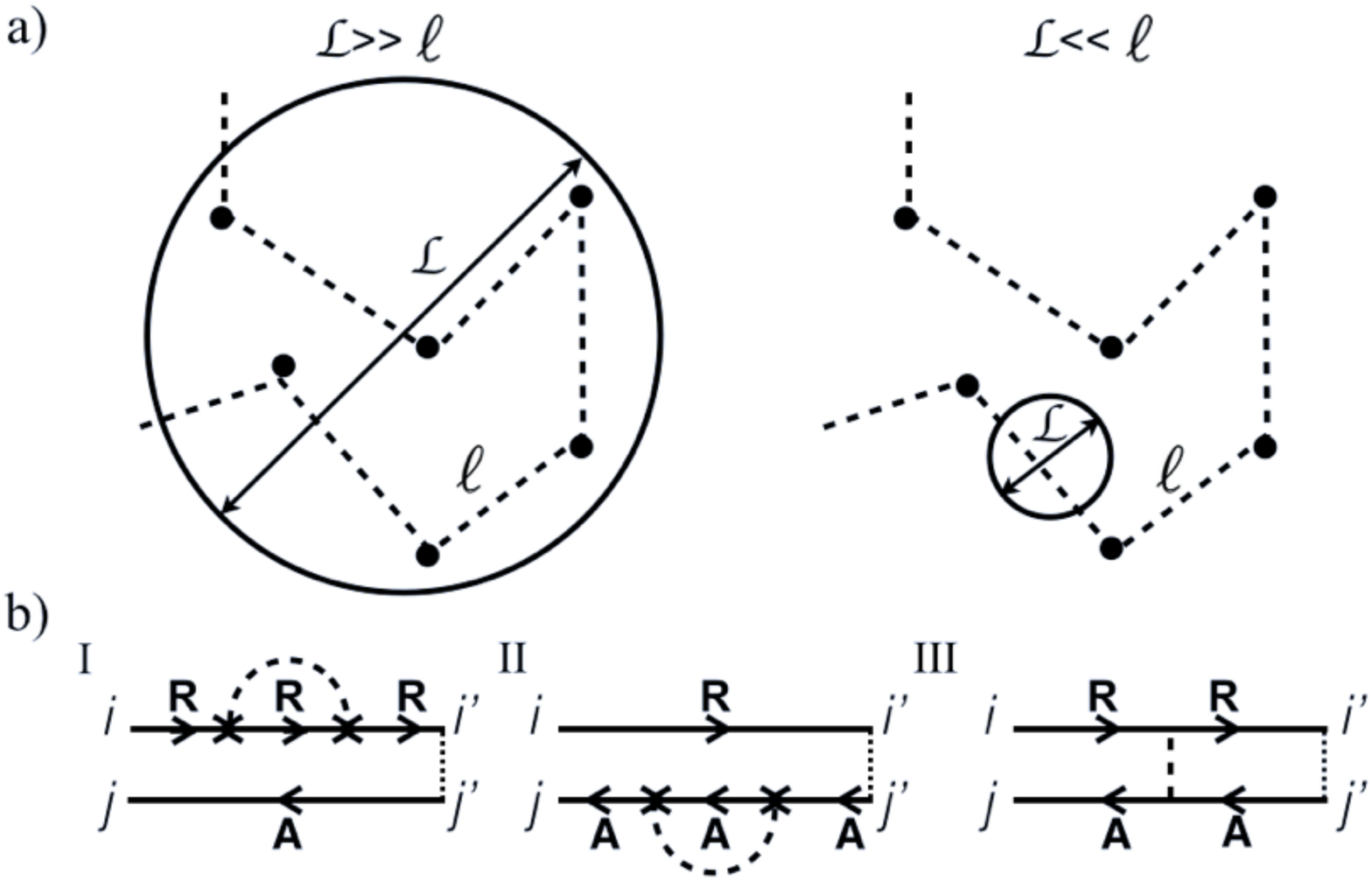}
\caption{ a) Scheme of the motional narrowing in the D'yakonov-Perel' (left) and fluctuations-dominated regimes (right). 
b) Diagrams corresponding to the second order correction to the diffusion pole $P_{iji'j'}$. The latin labels specify the spin projection with respect to the $z$-axis. Only the $z$-triplet mode $\frac{1}{2}\sum_{i,j,i',j'}[s_z]_{ji}P_{iji'j'}[s_z]_{i'j'}$ enters in Eq.~\eqref{eq:diff}.}
\label{fig:scheme}
\end{figure}

\textit{Spin diffusion limited by static ripples}.---We consider now a disordered graphene sample supported on a substrate, in which spin diffusion is assisted by motional narrowing. The competition between the 2 relevant length scales of the problem, namely, the electrons' mean free path, $\ell$, and the heights-correlation length imposed by the interaction with the substrate, $\mathcal{L}$, is illustrated in Fig.~\ref{fig:scheme}~a. The curvature of the sample is approximately uniform within a region of characteristic size $\mathcal{L}$. The electrons experience an effective exchange field that makes the spins to precess with a characteristic Larmor frequency of $\omega_L\sim \Delta_{\textrm{BR}}/\hbar$, where $\Delta_{BR}\sim\beta_{BR}\sqrt{\left\langle h^2\right\rangle}/\mathcal{L}^2$. The precession axis depends on the direction of motion, so momentum scattering randomizes the process when $\mathcal{L}>\ell$. In between scattering events, the electron spin precesses an angle $\phi\sim \tau\omega_L$, where $\tau=\ell/v_F$ is the scattering time. After a time $t$, and assuming that the process is Markovian, the precession angle is approximately $\phi\left(t\right)\sim\sqrt{t/\tau}\times\tau\omega_L$. On the contrary, if $\mathcal{L}\lesssim\ell$, then the precession is randomized by the fluctuations of the spin-orbit coupling itself. Within a region of size $\mathcal{L}$ the spin precess an angle $\phi\sim\mathcal{L}\,\omega_L/ v_F$, so after a time $t$ we have $\phi\left(t\right)\sim\sqrt{t v_F/\mathcal{L}}\times\mathcal{L}\,\omega_L/ v_F$. If we define the characteristic time scale of spin relaxation as $\phi\left(t=\tau_s\right)\sim1$, then from the previous arguments we obtain\begin{align}
\label{eq:asymptotic}
\frac{1}{\tau_s}\sim\begin{cases}
\frac{\tau \beta_{BR}^2\left\langle h^2\right\rangle}{\hbar^2\mathcal{L}^4} & \text{if}\,\,\mathcal{L}>\ell,\\
\frac{\beta_{BR}^2\left\langle h^2\right\rangle}{\hbar^2 v_F\mathcal{L}^3} & \text{otherwise}.
\end{cases}
\end{align}
The usual scaling $\tau_s^{-1}\propto\tau$ of the D'yakonov-Perel' mechanism saturates for scattering times larger than $\mathcal{L}/v_F$.

Next, we derive this qualitative result from a more rigorous diagrammatic calculation.\cite{diagrammatics} In the diffusive regime, the dynamics of the disorder-averaged spin density along the out-of-plane direction --$\rho_z\equiv\frac{1}{2}\text{Tr}\left[ s_z\hat{\rho}\right]$, where $\hat{\rho}$ is the density matrix operator-- is described by\begin{align}
\label{eq:diff}
\left(\partial_t-D\nabla^2+\frac{1}{\tau_s}\right)\rho_z=0,
\end{align}
where $D=v_F^2\tau/2$ is the diffusion constant. In the ladder approximation, the spin-relaxation rate $\tau_s^{-1}$ is given by the correction to the diffusion pole of the $z$-triplet mode of the 2-particle correlation function. This correction arises from the space-dependent spin-lattice coupling, which is treated in perturbation theory. The diagrams to the lowest order in $\beta_{BR}$ are shown in Fig.~\ref{fig:scheme}~b. The dashed lines correspond to the correlation function $\left\langle h\left(\mathbf{r}\right) h\left(\mathbf{r}'\right) \right\rangle$, the interaction vertex is the Bychkov-Rashba coupling, and the straight lines are disorder-averaged Green functions within the Born approximation,\begin{align}
\hat{G}^{R,A}\left(\omega,\mathbf{k}\right)=\left(\omega\pm\frac{i\hbar}{2\tau}-
\hbar v_F\boldsymbol{\Sigma}\cdot\mathbf{k}\right)^{-1}.
\label{eq:G}
\end{align}
The calculation is highly simplified if we neglect inter-band transitions leading to Elliot-Yafet-like contributions, which are expected to be parametrically small for usual dopings.\cite{EY2} The final result reads\begin{align}
\label{eq:result}
\frac{1}{\tau_s}=\frac{\beta_{BR}^2\tau}{\hbar^2}\int\frac{d^2\mathbf{q}}{\left(2\pi\right)^2}\,\frac{\left|\mathbf{q}\right|^4\left\langle 
\left|h\left(\mathbf{q}\right)\right|^2\right\rangle}{\sqrt{1+\tau^2v_F^2\left|\mathbf{q}\right|^2}},
\end{align}
where $\left\langle \left|h\left(\mathbf{q}\right)\right|^2\right\rangle$ is the correlation function in momentum space.

For simplicity, we may consider a correlation function of the form $\left\langle h\left(\mathbf{r}\right) h\left(0\right) \right\rangle=h_0^2\,e^{-\left|\mathbf{r}_1\right|^2/\mathcal{L}^2}$, where $h_0\approx0.3$ nm corresponds to the characteristic height of the ripples and the correlation length $\mathcal{L}\approx 25$ nm
is a measure of their typical lateral size.\cite{exp} By performing the Fourier transform and plugging the result into Eq.~\eqref{eq:result} we obtain\begin{align}
\label{eq:result_asymptotic}
&\frac{1}{\tau_s}=\frac{\beta_{BR}^2h_0^2}{2\hbar^2 v_F\mathcal{L}^3}\times\int_0^{\infty}d\zeta\text{ }
\frac{\zeta^5e^{-\zeta^2/4}}{\sqrt{\zeta^2+\mathcal{L}^2/\ell^2}}
\\
&\approx\begin{cases}
\frac{32\beta_{BR}^2h_0^2\tau}{\hbar^2\mathcal{L}^4}&\text{if }\mathcal{L}\gg\ell,\\
\frac{6\sqrt{\pi}\beta_{BR}^2h_0^2}{\hbar^2 v_F\mathcal{L}^3}&\text{if }\mathcal{L}\ll\ell.
\end{cases}
\nonumber
\end{align}
The results in the asymptotic regimes coincide with our estimates in Eq.~\eqref{eq:asymptotic} up to numerical factors.\cite{foot_diff} The spin lifetimes in the limit $\mathcal{L}\lesssim\ell$ are of the order of $\tau_s\sim10$ $\mu$s. This estimation only includes the spin-orbit coupling of carbon atoms. The substrate itself can enhance substantially the strength of the spin-orbit coupling, leading to much shorter spin-relaxation times.

\begin{figure}
\includegraphics[width=1\columnwidth]{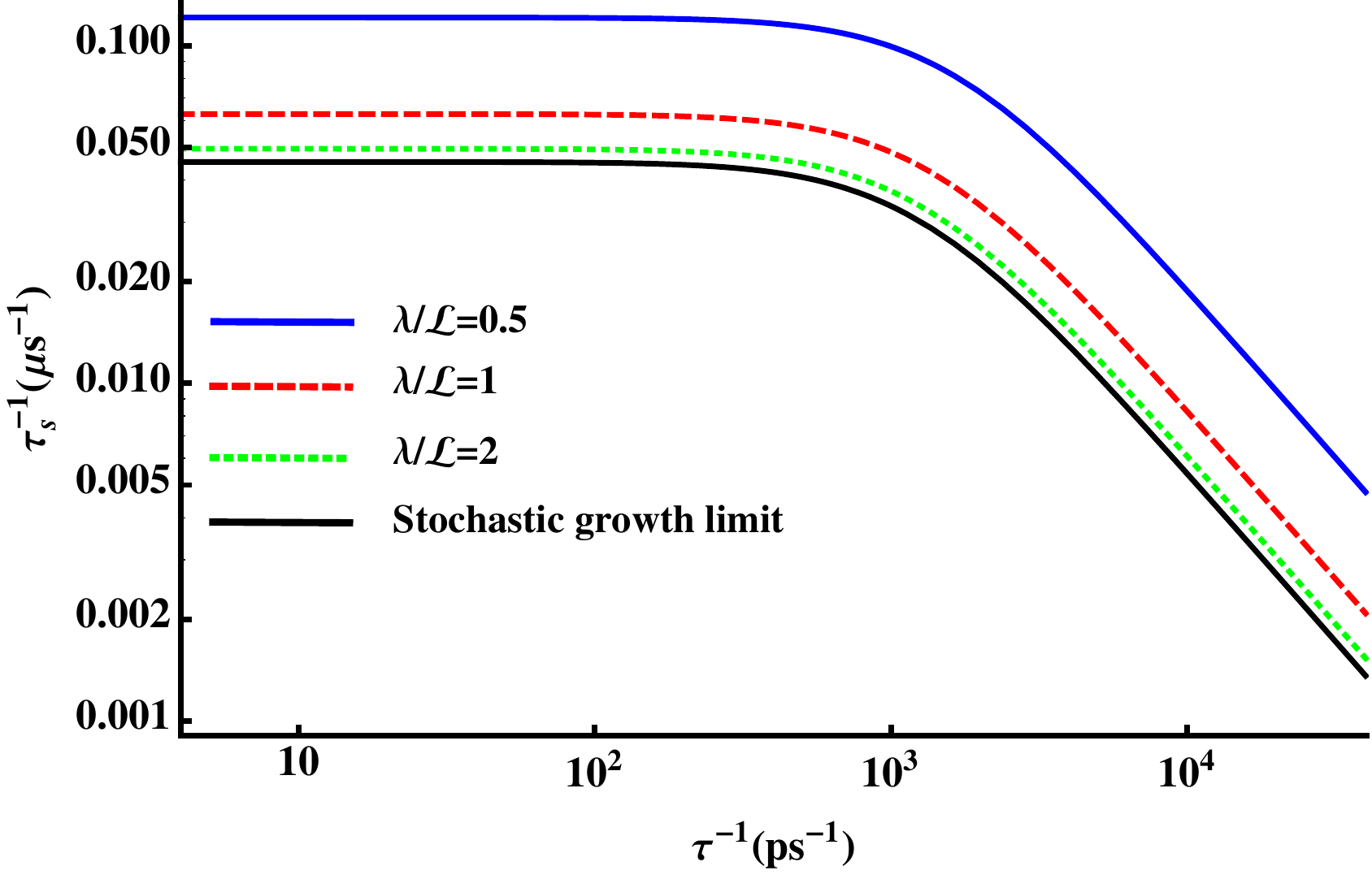}
\caption{Spin-relaxation rates as a function of $\tau^{-1}=v_F/\ell$ evaluated from Eqs.~\eqref{eq:result}-\eqref{eq:cor}. The result in Eq.~\eqref{eq:result_asymptotic} corresponds to the continuous black line.}
\label{fig:diffusive}
\end{figure}

\textit{Discussion}.---Our formula in Eq.~\eqref{eq:result} can be applied to the study of spin relaxation in epitaxial graphene.\cite{Fert_spin_injection,spin_epitaxial} The correlator used in the derivation of Eq.~\eqref{eq:result_asymptotic} describes a noise-induced roughening of the epitaxial growth fronts, which is a scale-invariant (self-affine) random process.\cite{34} In graphene samples, however, a preferential periodicity has been systematically observed,\cite{exp} described by a correlation function of the form\cite{34} 
\begin{align}
\label{eq:cor}
\left\langle h\left(\mathbf{r}\right) h\left(0\right) \right\rangle
=h_0^2\text{ }e^{-\frac{\left|\mathbf{r}_1\right|^2}{\mathcal{L}^2}}\mathcal{J}_0
\left(\frac{\left|\mathbf{r}_1-\mathbf{r}_2\right|}{\lambda}\right),
\end{align}
where $\mathcal{J}_0\left(x\right)$ is a Bessel function of the first kind. Figure~\ref{fig:diffusive} shows the spin-relaxation rate as a function of the inverse of the scattering time for different values of $\lambda$. There are still two asymptotic regimes dominated by momentum scattering (D'yakonov-Perel' mechanism) and height fluctuations, regardless the actual value of $\lambda$.

In summary, we have analyzed the role of lattice corrugations and thermal out-of-plane vibrations in the spin transport of graphene. Flexural phonons give rise to a temperature-dependent contribution to spin relaxation; for the usual carrier concentrations, the spin lifetimes are of the order of $0.1-1$ $\mu$s at room temperature, depending on the amount of strain in the sample and the interaction with the substrate. Static ripples also affect the spin transport in the diffusive regime. In the limit $\mathcal{L}\lesssim \ell$, the spin lifetimes are exclusively determined by the geometry of the corrugations. The subtraction of the effect of the contacts in the analysis of the Hanle-precession curves\cite{Vivek_Aji_contacts,Otani_Fert_contacts} makes possible to study these relaxation mechanisms in graphene-based spin valves.

We acknowledge financial support from MINECO (Spain), grant FIS2014-57432-P, ERC Advanced Grant NOVGRAPHENE (Nr. 290846), the 
European Commission under the Graphene Flagship, contract CNECTICT-604391, and the U.S. Department of Energy, Office of Basic Energy Sciences, Division of Materials Sciences and Engineering under contract DE-SC0012190 (H.O.).

\end{document}